# SPECTRAL FINGERPRINTS OF EARTH-LIKE PLANETS AROUND FGK STARS

SARAH RUGHEIMER,[1] LISA KALTENEGGER,[1,2] ANDRAS ZSOM,[2,3] ANTÍGONA SEGURA,[4] AND DIMITAR SASSELOV[1]

[1]Harvard Smithsonian Center for Astrophysics, 60 Garden st., 02138 MA Cambridge, USA
[2]MPIA, Koenigstuhl 17, 69117 Heidelberg, Germany
[3]Department of Earth, Atmospheric and Planetary Sciences, Massachusetts Institute of Technology, Cambridge, MA 02139
[4]Instituto de Ciencias Nucleares, Universidad Nacional Autónoma de México, México

**Please send editorial correspondence to**: Sarah Rugheimer, Center for Astrophysics, 60 Garden St. MS 10, Cambridge, MA 02138. srugheimer@cfa.harvard.edu. W: 617-496-0741 C: 406-871-7466. F: 617-495-7008

## ABSTRACT

We present model atmospheres for an Earth-like planet orbiting the entire grid of main sequence FGK stars with effective temperatures ranging from $T_{eff}$ = 4250K to $T_{eff}$ = 7000K in 250K intervals. We model the remotely detectable spectra of Earth-like planets for clear and cloudy atmospheres at the 1AU equivalent distance from the VIS to IR (0.4 μm - 20 μm) to compare detectability of features in different wavelength ranges in accordance with JWST and future design concepts to characterize exo-Earths. We also explore the effect of the stellar UV levels as well as spectral energy distribution on a terrestrial atmosphere concentrating on detectable atmospheric features that indicate habitability on Earth, namely: $H_2O$, $O_3$, $CH_4$, $N_2O$ and $CH_3Cl$.

The increase in UV dominates changes of $O_3$, OH, $CH_4$, $N_2O$ and $CH_3Cl$ whereas the increase in stellar temperature dominates changes in $H_2O$. The overall effect as stellar effective temperatures and corresponding UV increase, is a lower surface temperature of the planet due to a bigger part of the stellar flux being reflected at short wavelengths, as well as increased photolysis. Earth-like atmospheric models show more $O_3$ and OH but less stratospheric $CH_4$, $N_2O$, $CH_3Cl$ and tropospheric $H_2O$ (but more stratospheric $H_2O$) with increasing effective temperature of Main Sequence stars. The corresponding spectral features on the other hand show different detectability depending on the wavelength observed.

We concentrate on directly imaged planets here as framework to interpret future lightcurves, direct imaging and secondary eclipse measurements of atmospheres of terrestrial planets in the HZ at varying orbital positions.

## 1. INTRODUCTION

Over 830 extrasolar planets have been found to date with thousands more candidate planets awaiting confirmation from NASA's Kepler Mission. Several of these planets have been found in or near the circumstellar Habitable Zone (see e.g. Batalha *et al*., 2012; Borucki *et al*. 2011, Udry *et al*., 2007; Kaltenegger & Sasselov 2011) with masses and radii consistent with rocky planet models. Recent radial velocity results as well as Kepler demonstrate that small planets in the Habitable Zone (HZ) exist around solar type stars. Future mission concepts to characterize Earth-like planets are designed to take spectra of extrasolar planets with the ultimate goal of remotely detecting atmospheric signatures (e.g., Beichman *et al*., 1999, 2006; Cash 2006; Traub *et al.,* 2006). For transiting terrestrial planets around the closest stars, the James Web Space Telescope (JWST, see Gardner *et al*., 2006) as well as future ground and space based telescopes might be able to detect biosignatures by adding multiple transits for the closest stars (see discussion).

Several groups have explored the effect of stellar spectral types on the atmospheric composition of Earth-like planets by considering specific stars: F9V and K2V (Selsis, 2000); F2V and K2V (Segura *et al*., 2003; Grenfell *et al*., 2007; Kitzmann *et al*., 2011ab). In this paper we expand on this work by establishing planetary atmosphere models for the full FGK main sequence, using a stellar temperature grid from 7000K to 4250K, in increments of 250K, to explore the effect of the stellar types on terrestrial atmosphere models. We show the effects of stellar UV and stellar temperature on the planet's atmosphere individually to understand the overall effect of the stellar type on the remotely detectable planetary spectrum from 0.4-20 μm for clear and cloudy atmosphere models.. This stellar temperature grid covers the full FGK spectral range and corresponds roughly to F0V, F2V, F5V, F7V, F9V/G0V, G2V, G8V, K0V, K2V, K4V, K5V and K7V main sequence stars (following the spectral type classification by Gray, 1992).

In this paper we use "Earth-like", as applied to our models, to mean using modern Earths outgassing rates (following Segura et al. 2003). We explore the influence of stellar spectral energy distribution (SED) on the chemical abundance and planetary atmospheric spectral features for Earth-like planets including biosignatures and their observability from the VIS to IR. Atmospheric biosignatures are chemical species in the atmosphere that are out of chemical equilibrium or are byproducts of life processes. In our analysis we focus particularly on spectral features of chemical species that indicate habitability for a temperate rocky planet like Earth, H2O, O3, CH4, N2O and CH3Cl (Lovelock, 1975; Sagan *et al*., 1993).

In Section 1 we introduce the photochemistry of an Earth-like atmosphere. In Section 2, we describe our model for calculating the stellar spectra, atmospheric models, and planetary spectra. Section 3 presents the influence of stellar types on the abundance of various atmospheric chemical species. In Section 4 we examine the remote observability of such spectral features, and in Sections 5 and 6 we conclude by summarizing the results and discussing their implications.

*1.1 Photochemistry for Earth-like planets including potential biosignatures*

For an Earth-like biosphere, the main detectable atmospheric chemical signatures that in combination could indicate habitability are $O_2/O_3$ with $CH_4/N_2O$, and $CH_3Cl$. Note that one spectral feature e.g. $O_2$ does not constitute a biosignature by itself as the planetary context (like bulk planet, atmospheric composition and planet insolation) must be taken into account to interpret this signature. Detecting high concentrations of a reducing gas concurrently with $O_2$ or $O_3$ can be used as a biosignature since reduced gases and oxygen react rapidly with each other. Both being present in significant and therefore detectable amounts in low resolution spectra implies a strong source of both. In the IR, $O_3$ can be used as a proxy for oxygen at $10^{-2}$ Present Atmospheric Level of $O_2$, the depth of the 9.6 μm $O_3$ feature is comparable to the modern atmospheric level (Kasting *et al*., 1985; Segura *et al*., 2003). At the same time, because of the 9.6 μm $O_3$ feature's non-linear dependence on the $O_2$ concentration, observing in the visible at 0.76 μm would be a more accurate $O_2$ level indicator, but requires higher resolution than detecting $O_3$.

$N_2O$ and $CH_3Cl$ are both primarily produced by life on Earth with no strong abiotic sources, however, their spectral features are likely too small to detect in low resolution with the first generation of missions. While $H_2O$ or $CO_2$ are not considered biosignatures as both are produced through abiotic processes, they are important indicators of habitability as raw materials and can indicate the level of greenhouse effect on a planet. We refer the reader to other work (e.g. Des Marais *et al*., 2006; Meadows 2006; and Kaltenegger *et al*., 2010b) for a more in depth discussion on habitability and biosignatures. In this section we briefly discuss the most important photochemical reactions involving: $H_2O$, $O_2$, $O_3$, $CH_4$, $N_2O$, and $CH_3Cl$.

*Water, $H_2O$*: Water vapor is an important greenhouse gas in Earth's atmosphere. Over 99% of $H_2O$ vapor is currently in the troposphere, where it is an important source of OH via the following set of reactions:

$$O_3 + h\nu \rightarrow O_2 + O(^1D) \quad [R1]$$

$$H_2O + O(^1D) \rightarrow 2OH \quad [R2]$$

In the troposphere, the production of $O(^1D)$ takes place for 3000 Å $< \lambda <$ 3200 Å, the lower limit of which is set by the inability of wavelengths, λ shorter than 3000 Å to reach the troposphere due to $O_3$ shielding. $H_2O$, while photochemically inert in the troposphere, can be removed by photolysis primarily by wavelengths shortward of 2000 Å in the stratosphere. The photodissociation threshold energy is 2398 Å, but the cross-section of the molecule above 2000 Å is very low. Stratospheric $H_2O$ can be transported from the troposphere or be formed in the stratosphere by $CH_4$ and OH.

$$CH_4 + OH \rightarrow CH_3 + H_2O \quad [R3]$$

*Oxygen and Ozone, $O_2$ and $O_3$*: In an atmosphere containing $O_2$, $O_3$ concentrations are determined by the absorption of ultraviolet (UV) light shortward of 2400 Å in the stratosphere. $O_3$ is an oxidizing agent more reactive than $O_2$, the most stable form of oxygen, due to the third oxygen atom being loosely bound by a single bond. $O_3$ is also an indirect measure of OH since reactions involving $O_3$ and $H_2O$ are sources of OH. OH is very reactive and is the main sink for reducing species such as $CH_4$. $O_3$ is formed primarily by the Chapman reactions (1930) of the photolysis of $O_2$ by UV photons (1850 Å $< \lambda <$ 2420 Å) and then the combining of $O_2$ with O.

$$O_2 + h\nu \rightarrow O + O \quad (\lambda < 240 nm) \quad [R4]$$

$$O + O_2 + M \rightarrow O_3 + M \quad [R5]$$

$$O_3 + h\nu \rightarrow O_2 + O \quad (\lambda < 320 nm) \quad [R6]$$

$$O_3 + O \rightarrow 2O_2 \quad [R7]$$

where M is any background molecule such as $O_2$ or $N_2$. Reactions [R5] and [R6] are relatively fast compared with [R4] and [R7] which are the limiting reactions in Earth's atmosphere. However, considering the Chapman mechanism alone would overpredict the concentration of $O_3$ by a factor of two on Earth. Hydrogen oxide ($HO_x$), nitrogen oxide ($NO_x$), and chlorine ($ClO_x$) radicals are the additional sinks controlling the $O_3$ abundance (Bates & Nicolet, 1950; Crutzen, 1970; Molina and Rowland, 1974, respectively), with $NO_x$ and $HO_x$ being the dominant and second-most dominant sink, respectively.

*Methane, $CH_4$*: Since $CH_4$ is a reducing gas, it reacts with oxidizing species and thus has a short lifetime of around 10-12 years in modern Earth's atmosphere (Houghton *et al*. 2004). In both the troposphere and stratosphere, $CH_4$ is oxidized by OH, which is the largest sink of the global methane budget. In the stratosphere, $CH_4$ is also destroyed by UV radiation. Though its photodissociation energy is 2722 Å, its absorption cross-section isn't sufficient for λ > 1500 Å. $CH_4$ is produced biotically by methanogens and termites, and abiotically through hydrothermal vent systems. In the modern atmosphere there is a significant anthropogenic source of $CH_4$ from natural gas, livestock, and rice paddies. $CH_4$ is 25x more effective as a greenhouse gas than $CO_2$ in modern Earth's atmosphere (Forster *et al*., 2007) and may have been much more abundant in the early Earth (see e.g. Pavlov *et al*., 2003).

*Nitrous Oxide, $N_2O$*: Nitrous oxide, $N_2O$, is a relatively minor constituent of the modern atmosphere at around 320 ppbv, with a pre-industrial concentration of 270 ppbv (Forster *et al*., 2007). It is important for stratospheric chemistry since around 5% is converted to NO, an important sink of $O_3$, and 95% produces $N_2$.

$$N_2O + O(^1D) \rightarrow 2NO \quad [R8]$$

On current Earth, N$_2$O is emitted primarily by denitrifying bacteria with anthropogenic sources from fertilizers in agriculture, biomass burning, industry and livestock.

*Methyl Chloride, CH$_3$Cl*: CH$_3$Cl has been proposed as a potential biosignature because its primary sources are marine organisms, reactions of sea foam and light, and biomass burning (Segura *et al*., 2005). The primary loss of CH$_3$Cl in Earth's atmosphere is by OH as seen in [R9], but it can also be photolyzed or react with atomic chlorine. Because CH$_3$Cl is a source of chlorine in the stratosphere, it also plays a role in the removal of O$_3$ as discussed earlier.

$$CH_3Cl + OH \rightarrow Cl + H_2O \quad [R9]$$
$$CH_3Cl + h\nu \rightarrow CH_3 + Cl \quad [R10]$$
$$CH_3Cl + Cl \rightarrow HCl + Cl \quad [R11]$$

## 2. Model description

We use EXO-P (Kaltenegger & Sasselov 2010) a coupled one-dimensional radiative-convective atmosphere code developed for rocky exoplanets based on a 1D climate (Kasting & Ackerman 1986, Pavlov *et al*. 2000, Haqq-Misra *et al*. 2008), 1D photochemistry (Pavlov & Kasting 2002, Segura *et al*. 2005, 2007) and 1D radiative transfer model (Traub & Stier 1978, Kaltenegger & Traub 2009) to calculate the model spectrum of an Earth-like exoplanet.

*2.1 Planetary Atmosphere Model*

EXO-P is a model that simulates both the effects of stellar radiation on a planetary environment and the planet's outgoing spectrum. The altitude range extends to 60km with 100 layers. We use a geometrical model in which the average 1D global atmospheric model profile is generated using a plane parallel atmosphere, treating the planet as a Lambertian sphere, and setting the stellar zenith angle to 60 degrees to represent the average incoming stellar flux on the dayside of the planet (see also Schindler & Kasting, 2000).

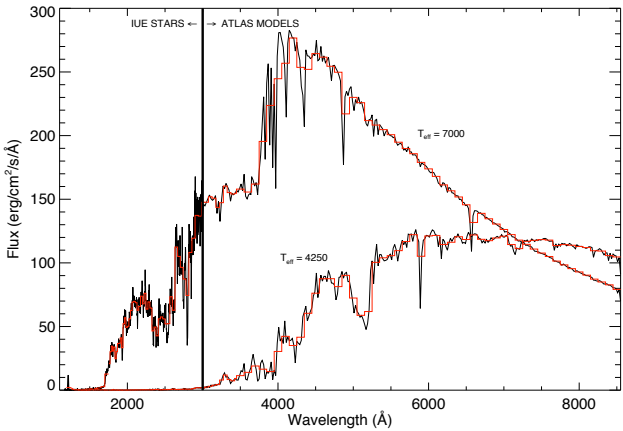

Figure 1: F0V and K7V composite input stellar spectrum of IUE observations coadded to (black) ATLAS photospheric models (Kurucz, 1979) and (red) binned stellar input. Note: the full input spectrum extends to 45450 Å. Only the hottest and coolest star in our grid are shown here for comparison.

The temperature in each layer is calculated from the difference between the incoming and outgoing flux and the heat capacity of the atmosphere in each layer. If the lapse rate of a given layer is larger than the adiabatic lapse rate, it is adjusted to the adiabat until the atmosphere reaches equilibrium. A two-stream approximation (see Toon *et al*., 1989), which includes multiple scattering by atmospheric gases, is used in the visible/near IR to calculate the shortwave fluxes. Four-term, correlated-*k* coefficients parameterize the absorption by O$_3$, H$_2$O, O$_2$, and CH$_4$ in wavelength intervals shown in Fig. 1 (Pavlov *et al.,* 2000). In the thermal IR region, a rapid radiative transfer model (RRTM) calculates the longwave fluxes. Clouds are not explicitly calculated. The effects of clouds on the temperature/pressure profile are included by adjusting the surface albedo of the Earth-Sun system to have a surface temperature of 288K (see Kasting et al., 1984; Pavlov *et al.* 2000; Segura *et al.,* 2003, 2005). The photochemistry code, originally developed by Kasting *et al.* (1985) solves for 55 chemical species linked by 220 reactions using a reverse-Euler method (see Segura *et al.,* 2010 and references therein).

The radiative transfer model used to compute planetary spectra is based on a model originally developed for trace gas retrieval in Earth's atmospheric spectra (Traub & Stier 1976) and further developed for exoplanet transmission and emergent spectra (Kaltenegger *et al*., 2007; Kaltenegger & Traub, 2009; Kaltenegger 2010; Kaltenegger *et al.* 2010a). In this paper we model Earth's reflected and thermal emission spectra using 21 of the most spectroscopically significant molecules (H$_2$O, O$_3$, O$_2$, CH$_4$, CO$_2$, OH, CH$_3$Cl, NO$_2$, N$_2$O, HNO$_3$, CO, H$_2$S, SO$_2$, H$_2$O$_2$, NO, ClO, HOCl, HO$_2$, H$_2$CO, N$_2$O$_5$, and HCl).

Using 34 layers the spectrum is calculated at high spectral resolution, with several points per line width, where the line shapes and widths are computed using Doppler and pressure broadening on a line-by-line basis, for each layer in the model atmosphere. The overall high-resolution spectrum is calculated with 0.1 cm$^{-1}$ wavenumber steps. The figures are shown smoothed to a resolving power of 250 in the IR and 800 in the VIS using a triangular smoothing kernel. The spectra may further be binned corresponding to proposed future spectroscopy missions designs to characterize Earth-like planets.

*2.2 Model Validation with EPOXI*

We previously validated EXO-P from the VIS to the infrared using data from ground and space (Kaltenegger *et al*., 2007). Here we use new data by EPOXI in the visible and near-infrared (Livengood *et al*., 2011) for further validation (see Fig. 2). The data set we use to validate our visible and the near-infrared Earth model spectra is the first EPOXI observation of Earth which was averaged over 24 hours on 03/18/2008 – 03/19/2008 and taken at a phase angle of 57.7°. The uncertainty in the EPOXI calibration is ~10% (Klassen *et al.,* 2008). Atmospheric models found the best match to be for a 50% cloud coverage with 1.5km and 8.5km cloud layer respectively (Robinson *et al*., 2011).

Here we use a 60% global cloud cover spectrum divided between three layers: 40% water clouds at 1km, 40% water clouds at 6km, and 20% ice clouds at 12km (following Kaltenegger *et al*., 2007) consistent with an averaged Earth profile to compare our model to this 24hr data set, which should introduce slight discrepancies. To correct the brightness values to match to our full-phase model we use a Lambert phase function.

Our model agrees with EPOXI on an absolute scale within 1-3% for the middle photometric points. The largest discrepancies in the visible are at 0.45 μm and 0.95 μm (with a 8% and 18% error respectively).

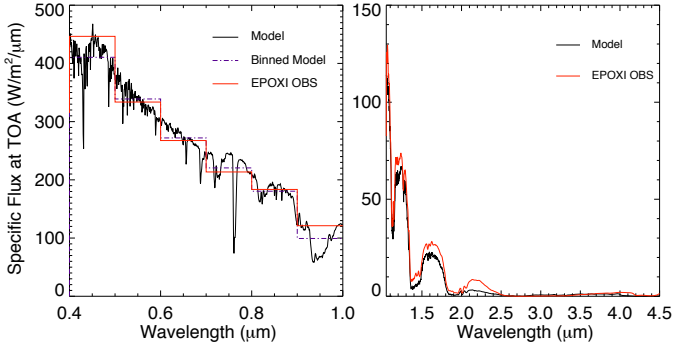

Figure 2: Comparison of EPOXI data (red) with the Earth model, top-of-atmosphere spectrum at full phase from EXO-P (black) in the visible (left) and near-infrared (right).

*2.3 Stellar Spectral Grid Model*

The stellar spectra grid ranges from 4250K to 7000K in effective temperature increments of 250K. This temperature range effectively probes the F0 to K7 main sequence spectral types. For each model star on our grid we concatenated a solar metallicity, unreddened synthetic ATLAS spectrum, which only considers photospheric emission (Kurucz, 1979), with observations from the International Ultraviolet Explorer (IUE) archive.[1] We use IUE measurements to extend ATLAS synthetic spectra, to generate input spectra files from 1150Å to 45,450Å (see Figs. 1 and 2). We choose main sequence stars in the IUE archive with corresponding temperatures close to the grid temperatures and near solar metallicity, as described below.

| Star | $T_{eff}$(K) | $T_{eff}$(K) Grid | [Fe/H] | Spectral Type Grid |
|---|---|---|---|---|
| η Lep | 7060 | 7000 | -0.13 | F0V |
| σ Boo | 6730 | 6750 | -0.43 | F2V |
| π³ Ori | 6450 | 6500 | 0.03 | F5V |
| ι Psc | 6240 | 6250 | -0.09 | F7V |
| β Com | 5960 | 6000 | 0.07 | F9V/G0V |
| α Cen A | 5770 | 5750 | 0.21 | G2V |
| τ Ceti | 5500 | 5500 | -0.52 | G8V |
| HD 10780 | 5260 | 5250 | 0.03 | K0V |
| ε Eri | 5090 | 5000 | -0.03 | K2V |
| ε Indi | 4730 | 4750 | -0.23 | K4V |
| 61 Cyg A | 4500 | 4500 | -0.43 | K5V |
| BY Dra | 4200 | 4250 | 0.00 | K7V |

Table 1: List of representative IUE stars with their measured $T_{eff}$, the $T_{eff}$ which corresponds to our grid of stars, their metallicity, and their approximate stellar type following Gray (1992).

The IUE satellite had three main cameras, the longwave (LWP/LWR) cameras (1850Å – 3350Å), and the shortwave (SW) camera (1150Å – 1975Å). When preparing the IUE data (following Segura *et al*., 2003; Massa *et al*., 1998; Massa &

---
[1] http://archive.stsci.edu/iue/

Fitzpatrick, 2000), we used a sigma-weighted average to coadd the multiple SW and LW observations. We used a linear interpolation when there was insufficient high quality measurements to merge the wavelength region from the SW to the LW cameras. IUE measurements were joined to ATLAS model spectra at 3000 Å. In a few cases, a shift factor is needed to match the IUE data to the ATLAS model (see also Segura *et al*., 2003) but unless stated explicitly no shift factor was used. Effective temperatures and metallicities are taken from NStED (derived from Flower *et al*. (1996) and Valenti & Fischer (2005), respectively) unless otherwise cited. See Table 1 for a summary list of the representative IUE stars chosen.

HD 40136, η Lep, is at 15.04pc with $T_{eff}$ = 7060K and [Fe/H] = -0.13 (Cayrel de Strobel *et al*., 2001), corresponding to an F0V, the hottest model grid star. Two LW and four SW spectra were coadded and merged with a 7000K ATLAS spectrum.

To compare with previous work (Segura *et al*., 2003; Grenfell *et al.,* 2007; Selsis, 2000), we chose HD 128167, σ Boötis, for our model F2V grid star. σ Boötis is an F2V star at 15.47pc with $T_{eff}$ = 6730K and [Fe/H] = -0.43. Two LW and five SW spectra were coadded and merged with a 6750K ATLAS spectrum. A slight downward shift of a factor of 0.88 is necessary to match the IUE data with a ATLAS spectrum (see also Segura *et al.,* 2003).

π³ Orionis, HD 30652, is at 8.03pc with a $T_{eff}$ = 6450K and [Fe/H] = 0.03, corresponding to an F5V grid star. Two LW and three SW spectra were coadded and merged with a 6500K ATLAS spectrum.

ι Piscium, HD 222368, is at 13.79pc with a $T_{eff}$ = 6240K and [Fe/H]= -0.09, corresponding to an F7V grid star. Two LW and four SW spectra were coadded and merged with a 6250K ATLAS spectrum.

β Com, HD 114710, is at 9.15pc with a $T_{eff}$ = 5960K and [Fe/H]= 0.07, corresponding to an G0V grid star. Only one LW spectrum was correctable with the Massa routines and thus one LW and five SW spectra were coadded and merged with a 6000K ATLAS spectrum.

α Centauri A, HD 128620, is at 1.35 pc with a $T_{eff}$ = 5770K and [Fe/H]= 0.21, corresponding to a G2V grid star. Three LW and 93 SW spectra were coadded and merged with an upward shift of 1.25 to a 5750K ATLAS spectrum.

τ Ceti, HD 10700, is at 3.65pc with $T_{eff}$ = 5500K and [Fe/H]=-0.52, corresponding to a G8V grid star. Two LW and eight SW spectra were coadded and merged with a 5500K ATLAS spectrum.

HD 10780 is at 9.98pc with $T_{eff}$ = 5260K and [Fe/H] = 0.03, corresponding to a K0V grid star. It is a variable of the BY Draconis type. Five LW and four SW spectra were coadded and merged with a 5250K ATLAS spectrum.

ε Eridani, HD 22049, is at 3.22pc with $T_{eff}$ = 5090K and [Fe/H] = -0.03, corresponding to a K2V gird star. ε Eri was chosen to compare with previous work (Segura *et al.,* 2003; Grenfell *et al.* 2007; Selsis, 2000). ε Eri is a young star, only 0.7 Ga (Di Folco *et al.,* 2004), and is thus more active than a typical K-dwarf. Due to its variability and close proximity there are frequent IUE observations. 17 LW and 72 SW IUE spectra were coadded and merged these with a 5000K ATLAS spectrum.

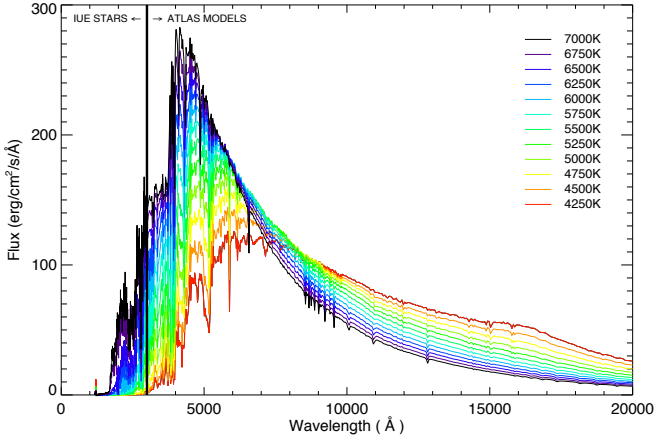

Figure 3: Composite stellar input spectra from IUE observations merged to a ATLAS photosphere model at 3000 Å for each grid star. We display up to 20,000 Å here however the complete input files extend to 45450 Å.

ε Indi, HD 209100, is at 3.63 pc with $T_{eff}$ = 4730K and [Fe/H] = -0.23, corresponding to a K4V grid star. Seven LW and 30 SW IUE spectra were coadded and merged with a 4750K ATLAS spectrum.

61 Cyg A, HD 201091, is at 3.48 pc with $T_{eff}$ = 4500K and [Fe/H] = -0.43 (Cayrel de Strobel et al., 2001), corresponding to a K5V grid star. 61 Cyg A is a variable star of the BY Draconis type. Six LW and twelve SW spectra were coadded and merged with an upward shift of 1.15 to match the 4500K ATLAS spectrum.

BY Dra, HD 234677, is at 16.42pc with a $T_{eff}$ = 4200K (Hartmann et al., 1977) and [Fe/H] = 0 (Cayrel de Strobel et al., 1997), corresponding to a K7V grid star. It is a variable of the BY Draconis type. Eight LW and 30 SW spectra were coadded and merged to the 4250K ATLAS spectrum.

All input stellar spectra are shown in Fig. 3.

*2.4 Simulation Set-Up*

To examine the effect of the SED of the host star on an Earth-like atmosphere, we build a temperature grid of stellar models ranging from 7000K to 4250K in steps of 250K, corresponding to F type stars to K dwarfs. We simulated an Earth-like planet with the same mass as Earth at the 1AU equivalent orbital distance, where the wavelength integrated stellar flux received on top of the planet's atmosphere is equivalent to 1AU in our solar system, 1370 Wm$^{-2}$.

The biogenic fluxes were held fixed in the models in accordance with the fluxes that reproduce the modern mixing ratios in the Earth-Sun case (following Segura et al., 2003). We first calculate the surface fluxes for long-lived gases $H_2$, $CH_4$, $N_2O$, CO and $CH_3Cl$. Simulating the Earth around the Sun with 100 layers yields a $T_{surf}$ = 288K for surface mixing ratios: $c_{H2}$ = 5.5 x 10$^{-7}$, $c_{CH4}$ = 1.6 x 10$^{-6}$, $c_{CO2}$ = 3.5 x 10$^{-4}$, $c_{N2O}$ = 3.0 x 10$^{-7}$, $c_{CO}$ = 9.0 x 10$^{-8}$, and $c_{CH3Cl}$ = 5.0 x 10$^{-10}$. The corresponding surface fluxes are -1.9 x 10$^{12}$ g $H_2$/year, 5.3 x 10$^{14}$ g $CH_4$/year, 7.9 x 10$^{12}$ g $N_2O$ per year, 1.8 x 10$^{15}$ g CO/year, and 4.3 x 10$^{12}$ g $CH_3Cl$/year. The best estimate for the modern $CH_4$ flux is 5.35 x 10$^{14}$ g/year (Houghton et al., 2004) and corresponds to the value derived in the model. Fluxes for the other biogenic species are poorly constrained. The $N_2$ concentration is set by the total surface pressure of 1 bar. To explore the effect of UV and temperature separately, we combine a certain ATLAS model with varying UV files and vise versa.

**3. ATMOSPHERIC MODEL RESULTS AND DISCUSSION**

The stellar spectrum has two effects on the atmosphere: first, the *UV effect* (§3.1) that primarily influences photochemistry and second, the *temperature effect* (§3.2) resulting from the difference in absorbed flux as a function of stellar SED. The same planet has a higher Bond albedo around hotter stars with SEDs peaking at shorter λ, where Rayleigh scattering is more efficient, than around cooler stars, assuming the same total stellar flux (Sneep & Ubachs, 2004). The overall resulting planetary Bond albedo that includes both atmospheric as well as surface albedo is calculated by the climate/photochemistry model and varies between 0.13 – 0.22 for planets around F0 stars to K7 stars respectively because of the stars' SED. Note that these values are lower than Earth's planetary Bond albedo of 0.31 because the warming effect of clouds is folded into the albedo value in the climate code, decreasing it artificially.

*3.1 The influence of UV levels on Earth-like atmosphere models (UV effect)*

To explore the effects of UV flux alone on the atmospheric abundance of different molecules, we combined specific IUE data files for stars with $T_{eff}$ = 7000K, 6000K and 4500K (representing high, mid and low UV flux) with a fixed ATLAS photospheric models of $T_{eff}$ = 6000K. The temperature/pressure and chemical profiles of this test are shown in panels a) of Figs. 4 and 5. Hot stars provide high UV flux in the 2000 – 3200 Å range, e.g. a F0V grid star emits 130x more flux in this wavelength range than a K7V grid star (Figs. 1 and 2).

| $T_{eff}$(K) Grid | Spectral Type Grid | Surface Temperature (K) | Ozone Column Depth (cm$^{-2}$) |
|---|---|---|---|
| 7000 | F0V | 279.9 | 1.2×10$^{19}$ |
| 6750 | F2V | 281.7 | 1.1×10$^{19}$ |
| 6500 | F5V | 283.2 | 9.6×10$^{18}$ |
| 6250 | F7V | 284.6 | 8.3×10$^{18}$ |
| 6000 | F9V/G0V | 286.4 | 7.3×10$^{18}$ |
| SUN | G2V | 288.1 | 5.3×10$^{18}$ |
| 5750 | G2V | 287.7 | 5.1×10$^{18}$ |
| 5500 | G8V | 289.1 | 3.2×10$^{18}$ |
| 5250 | K0V | 290.9 | 4.1×10$^{18}$ |
| 5000 | K2V | 291.9 | 3.3×10$^{18}$ |
| 4750 | K4V | 292.8 | 2.6×10$^{18}$ |
| 4500 | K5V | 297.0 | 2.6×10$^{18}$ |
| 4250 | K7V | 300.0 | 3.5×10$^{18}$ |

Table 2: Surface temperature and $O_3$ column depth for an Earth-like planet model orbiting the grid stars.

The Chapman reactions are driven primarily by photolysis in this wavelength range and the atmosphere models show an according increase in $O_3$ concentration and subsequent strong temperature inversion for planets orbiting hot grid stars (Table 2). The maximum heating in the stratosphere is a few kilometers above the peak of the $O_3$ concentration where both a high enough concentration of $O_3$ and a high enough flux of photons is present. $O_3$ abundance increases OH abundance, the primary sink of $CH_4$ and $CH_3Cl$. Figs. 5 and 7 shows a corresponding decrease in those molecules for high UV environment. $O_3$ shields $H_2O$ in the troposphere from UV environments. Stratospheric $H_2O$ is photolyzed by $\lambda < 2000$ Å or reacts with excited oxygen, $O^1D$ to produce OH radicals. Accordingly stratospheric $H_2O$ concentration decreases with decreasing UV flux. $N_2O$ decreases with increasing UV flux because of photolysis by $\lambda < 2200$ Å. $N_2O$ is also an indirect sink for stratospheric $O_3$ when it is converted to NO. Therefore decreasing $N_2O$ increases $O_3$ abundance. $O_2$ and $CO_2$ concentrations remain constant and well mixed for all stellar types.

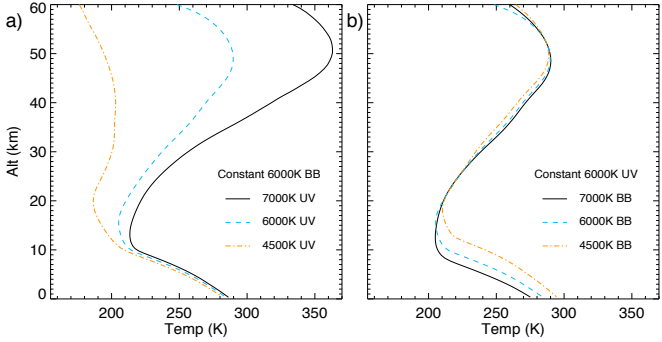

Figure 4. Temperature/altitude profiles for several unphysical test where we: a) combine high, mid, and low UV fluxes (IUE observations for stars with $T_{eff}$ = 7000K, 6000K, and 4500K, respectively) with a fixed ATLAS photosphere model for $T_{eff}$ = 6000K to show the "UV effect", and b) combine high, mid, and low stellar photosphere models (ATLAS models for $T_{eff}$ = 7000K, 6000K, and 4500K, respectively) with a fixed UV flux for $T_{eff}$ = 6000K to show the "Temperature effect".

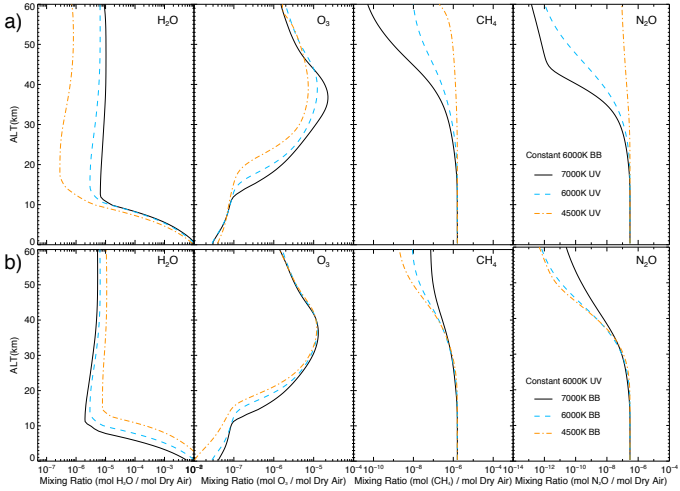

Figure 5. Chemical mixing ratio profiles for $H_2O$, $O_3$, $CH_4$, and $N_2O$ from several unphysical test where we: a) combine high, mid, and low UV fluxes (IUE observations for stars with $T_{eff}$ = 7000K, 6000K, and 4500K, respectively) with a fixed ATLAS photosphere model for $T_{eff}$ = 6000K to show the "UV effect", and b) combine high, mid, and low stellar photosphere models (ATLAS models for $T_{eff}$ = 7000K, 6000K, and 4500K, respectively) with a fixed UV flux for $T_{eff}$ = 6000K to show the "Temperature effect".

### 3.2 The influence of stellar $T_{eff}$ on Earth-like atmosphere models (Temperature effect)

To explore the effects of stellar $T_{eff}$ alone on the atmospheric abundance of different molecules, we combined specific photospheric ATLAS spectrum of $T_{eff}$ = 7000K, 6000K and 4500K (representing high, mid and low stellar $T_{eff}$) with a fixed UV data file of $T_{eff}$ = 6000K. The temperature/pressure and chemical profiles of this test are shown in panels b) of Figs. 4 and 5. $T_{eff}$ affects $H_2O$ vapor concentrations due to increased evaporation for high planetary surface temperature which is transported to the stratosphere. Fig. 4 shows an overall increase in tropopause and stratopause height for low stellar $T_{eff}$ with according hot planetary surface temperatures.

The response of $O_3$ to stellar $T_{eff}$ is weak due to two opposing effects: high stellar $T_{eff}$ and according low planetary surface and atmospheric temperatures increase $O_3$ concentration by slowing Chapman reactions that destroy $O_3$, but also increase $NO_x$, $HO_x$, and $ClO_x$ concentrations which are the primary sinks of $O_3$ (see also Grenfell et al., 2007).

Both $CH_4$ and $CH_3Cl$ show only a weak temperature dependence. The rate of the primary reactions of $CH_4$ and $CH_3Cl$ with OH slows with decreasing temperature, causing an increase in $CH_4$ and $CH_3Cl$ for lower planetary surface temperatures. $N_2O$ displays a similar weak temperature effect.

All of our simulations used a fixed mixing ratio of 355ppm for $CO_2$ and 21% $O_2$. Since both $O_2$ and $CO_2$ are well mixed in the atmosphere, their vertical mixing ratio profiles are not shown.

### 3.3 The influence of stellar SED on Earth-like atmosphere models

Figs. 6 and 7 show the combined temperature and UV effect on Earth-like atmospheres. The surface temperature of an Earth-like planet increases with decreasing stellar effective temperature due to decreasing reflected stellar radiation and increasing IR absorption by $H_2O$ and $CO_2$ (see Table 2 and Fig. 6). The late K-dwarf stars show in addition a near isothermal stratosphere.

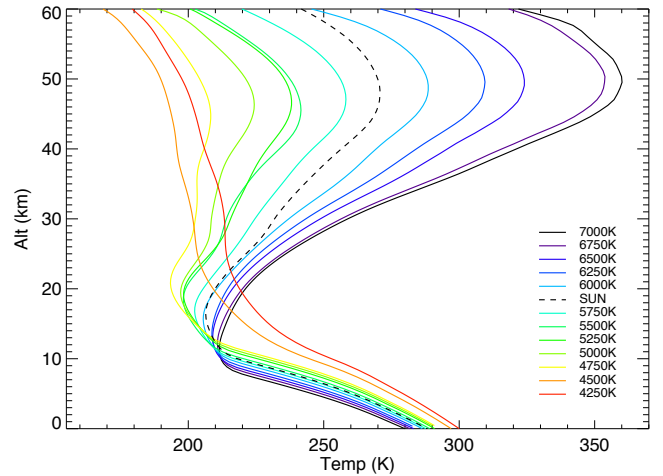

Figure 6. Planetary temperature/altitude profiles for different stellar types showing the combined temperature and UV effect.

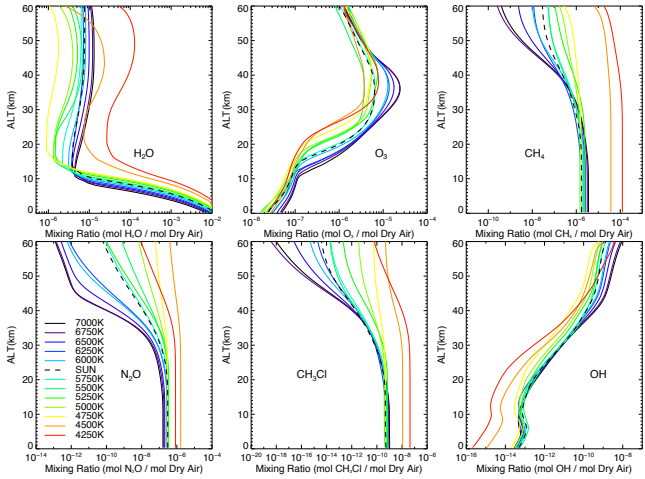

Figure 7: Photochemical model results for the mixing ratios of the major molecules $H_2O$, $O_3$, $CH_4$, $N_2O$, $CH_3Cl$, and OH for each stellar spectral type in our grid of stars showing the combined temperature and UV effect.

Fig. 7 shows the corresponding atmospheric mixing ratios versus height for the grid stars. The top height considered in our atmosphere models is 60 km for a Sun-like star, which corresponds to $10^{-4}$ bar (following Segura et al., 2003). For hotter stars the stratosphere is warmer increasing the pressure at 60 km to $4.0 \times 10^{-4}$ bars while for cooler stars the pressure at 60 km decreases to $3.0 \times 10^{-5}$ bars.

Earth-like atmosphere models around hot grid stars show high $O_3$ concentration (see Table 2) and therefore strong temperature inversions due to the increased stellar UV flux (Fig. 6). Cooler stars often have stronger emission lines and higher activity. Accordingly the coldest two grid stars in our sample ($T_{eff}$ = 4250K and 4500K) show a large $O_3$ abundance due to high stellar Ly α flux. In fact, the UV output of the coldest grid star, $T_{eff}$ = 4250K is almost 2x the UV flux of the second coldest grid star, $T_{eff}$ = 4500K, also due to its younger age. Thus, there is more $O_3$ produced for the coldest star. However, in the 2000 – 3000 Å wavelength region these cold grid stars emit low UV flux and therefore produce near isothermal stratospheres (see also M-dwarf models in Segura et al., 2005). The detailed effect of Ly α flux on the planet's atmosphere, will be modeled in a future paper.

Earth-like atmosphere models around hot grid stars also show high OH concentrations due to a higher availability of high energy photons, as well as $O_3$ and $H_2O$ molecules (Fig. 7). Cold grid stars ($T_{eff}$ = 4250K) show higher OH concentration in the stratosphere than expected from an extrapolation from the other grid stars due to the increased $O_3$ and $H_2O$ concentrations at those altitudes.

$CH_4$ abundance increases with decreasing stellar temperature, dominated by the effects of decreasing stellar UV. Stratospheric $CH_4$ decreases in atmosphere models around hot grid stars since both OH concentration and UV flux increase with stellar $T_{eff}$ and act as sinks of $CH_4$.

$H_2O$ abundance in the troposphere is dominated by the surface temperature of the planet. Earth-like planet atmosphere models around cool grid stars, generate warmer planetary surface temperatures, and therefore high amounts of tropospheric $H_2O$. High UV flux generally decreases $H_2O$ concentration in the stratosphere through photolysis but increased $O_3$ concentrations provides shielding from the photolysis of $H_2O$. Also cold grid stars ($T_{eff}$ = 4250K and 4500K) show increased stratosphere $H_2O$ concentration through increased vertical transport in the nearly isothermal stratospheres as well as production by stratospheric $CH_4$ (see e.g. Segura et al. 2005 for similar behavior in planets around M-dwarfs). In particular, the atmosphere models for a planet around $T_{eff}$ = 4250K grid star has a high OH concentration in the stratosphere due to increased $O_3$ and $H_2O$ at those altitudes.

$N_2O$ is primarily produced by denitrifying bacteria and has increased linearly due to agriculture since the preindustrial era at a rate of around 0.26% $yr^{-1}$ (Forster et al., 2007). Up to about 20km, there is no significant difference between stellar types in $N_2O$ concentration. Above ~20km, Fig. 7 shows a decrease in $N_2O$ concentration for atmosphere models around hot compared to cool grid stars since UV is the primary sink of $N_2O$ in the stratosphere. Below 20km $N_2O$ is shielded from photolysis by the $O_3$ layer. Note that the general trend for increasing $N_2O$ for colder grid stars reverses for our coldest grid star. This is due to the increased UV flux which destroys $N_2O$ and an increase in $O_3$ which causes an increase in $O(^1D)$, another strong sink for $N_2O$.

$CH_3Cl$ concentration decreases with increased stellar UV flux since OH which act as sink for $CH_3Cl$.

## 4. RESULTS: SPECTRA OF EARTH-LIKE PLANETS ORBITING F0V TO K7V GRID STARS

We include both a clear sky as well as a 60% global cloud cover spectrum which has cloud layers analogous to Earth (40% 1km, 40% 6km and 20% 12km following Kaltenegger et al., 2007) in Figs. 8-11 to show the importance of clouds on the reflected and emission planet spectra. We present the spectra as specific flux at the top of the atmosphere of Earth-like planets. In the VIS, the depth of the absorption features is primarily sensitive to the abundance of the species, while in the IR, both the abundance and the temperature difference between the emitting/absorbing layer and the continuum influences the depth of features.

We use a Lambert sphere as an approximation for the disk integrated planet in our model. The surface of our model planet corresponds to Earth's current surface of 70% ocean, 2% coast, and 28% land. The land surface consists of 30% grass, 30% trees, 9% granite, 9% basalt, 15% snow, and 7% sand. Surface reflectivities are taken from the USGS Digital Spectral Library[2] and the ASTER Spectral Library[3] (following Kaltenegger et al., 2007). Note the vegetation red edge feature at 0.76 μm is only detectable in the clear sky model spectra in low resolution, see Fig. 8 (see e.g Kaltenegger et al. 2007, Seager et al. 2002, Palle et al. 2008). No noise has been added to these model spectra to provide input models for a wide variety of instrument simulators for both secondary eclipse and direct detection simulations.

We assume full phase (secondary eclipse) for all spectra presented to show the maximum flux that can be observed.

---

[2] http://speclab.cr.usgs.gov/spectral-lib.html
[3] http://speclib.jpl.nasa.gov

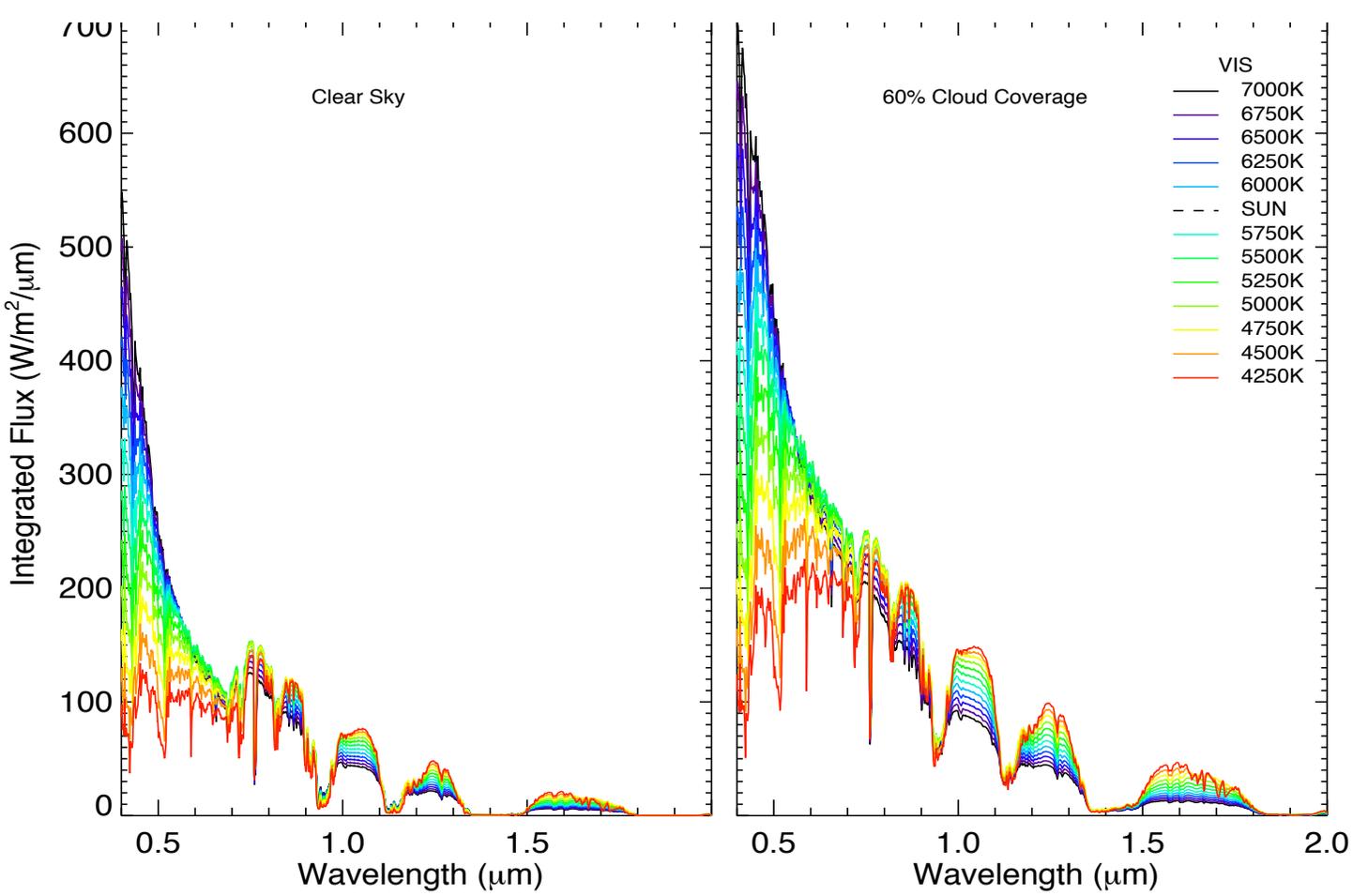

Figure 8: Smoothed, disk-integrated VIS/NIR spectra at the top of the atmosphere (TOA) for an Earth-like planet around FGK stars for both a clear sky (left) and 60% cloud coverage (right) model (region 2-4 μm has low integrated flux levels and therefore is not shown here).

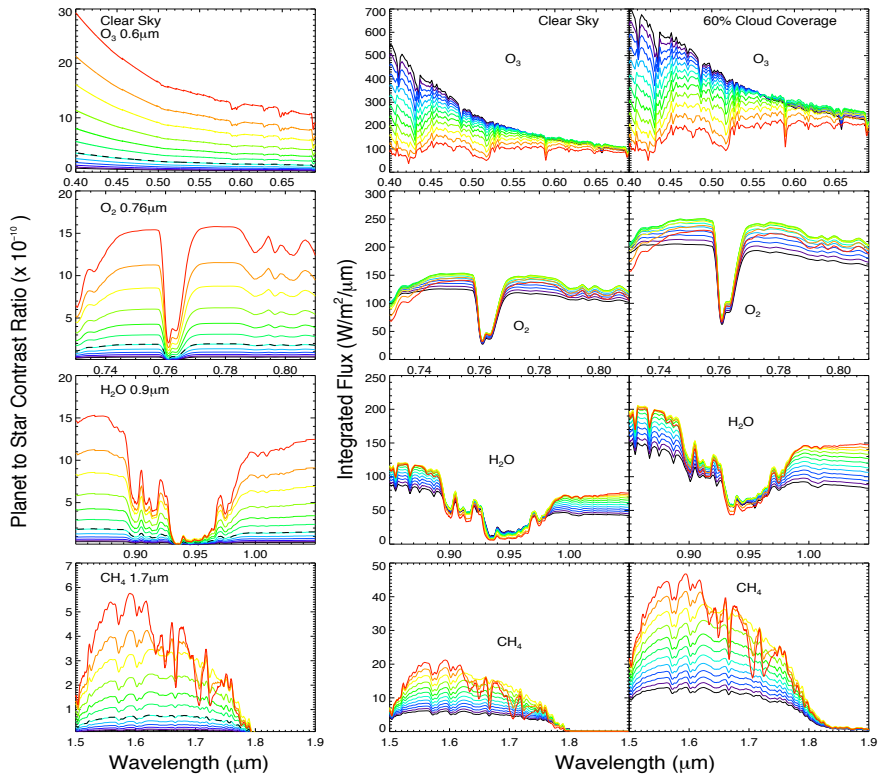

Figure 9: Individual features of $O_3$ at 0.6 μm, $O_2$ and 0.76 μm, $H_2O$ at 0.95 μm, and $CH_4$ at 1.7μm for F0V – K7V grid stars (left) planet-to-star contrast ratio and absolute flux levels (middle) for a clear sky and (right) 60% cloud coverage model. Note the different y-axes. Legend and color coding are the same in figures 6 to 11.

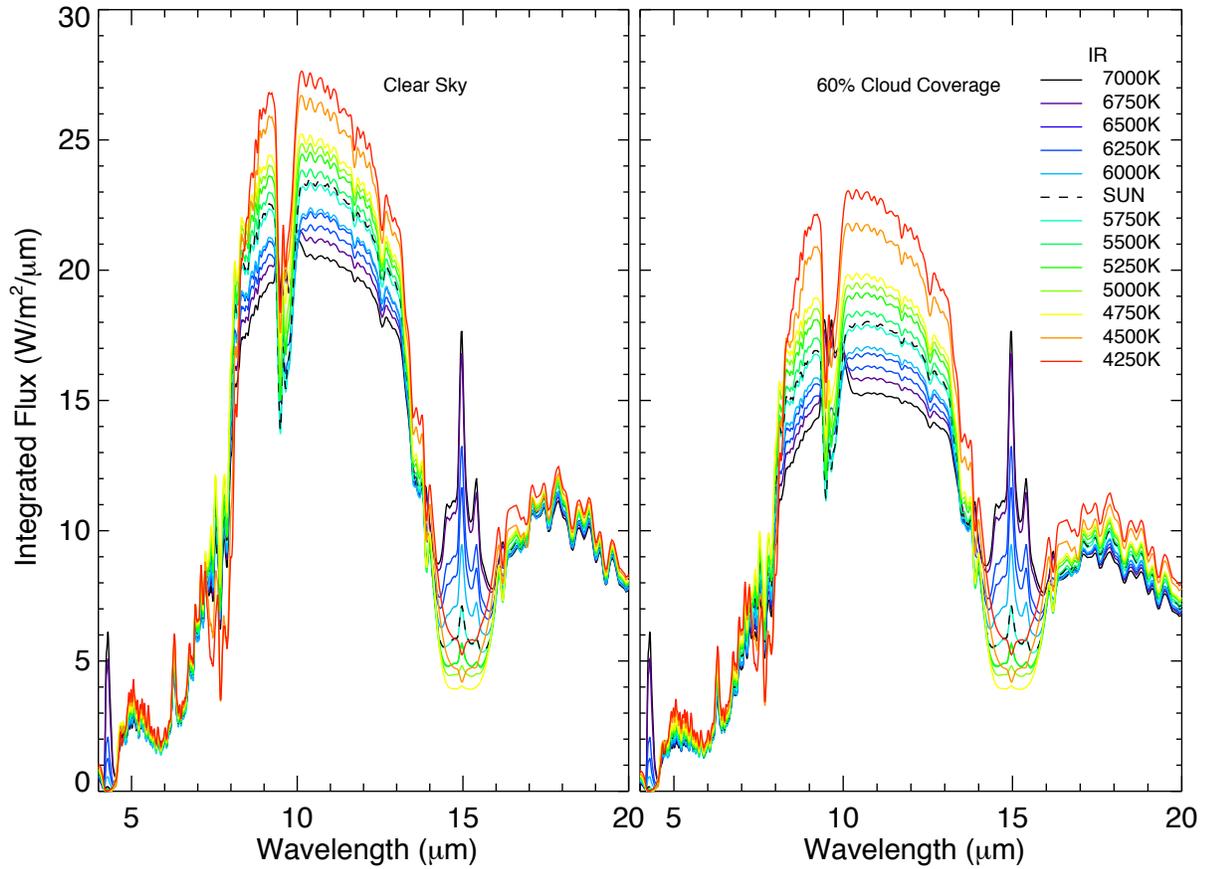

Figure 10: Smoothed, disk-integrated IR spectra at the top of the atmosphere (TOA) for Earth-like planets around F0V to K7V grid stars for both a clear sky (left) and 60% cloud coverage (right) model.

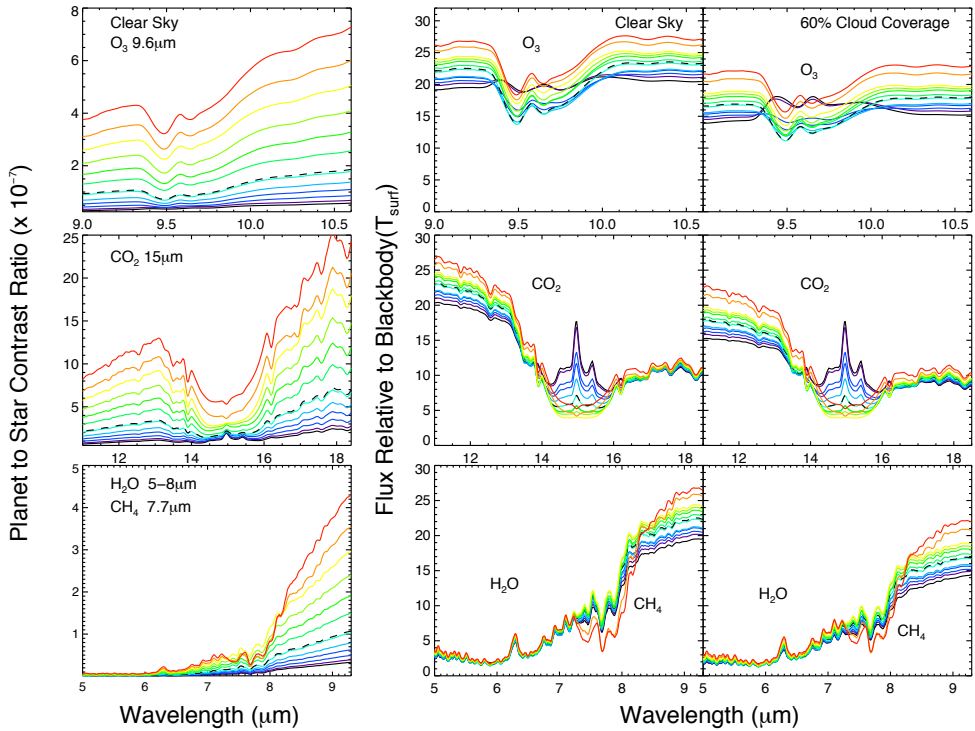

Figure 11: Individual features of $O_3$ at 9.6 μm, $CO_2$ and 15 μm, $H_2O$ at 5-8 μm, and $CH_4$ at 7.7μm for F0V – K7V grid stars (left) planet-to-star contrast ratio and absolute flux levels (middle) for a clear sky and (right) 60% cloud coverage model. Legend and color coding are the same in figures 2 to 8.

Note that we use an Earth-size planet to determine the specific flux and planet-to-star contrast ratio. A Super-Earth with up to twice Earth's radius will provide 4 times more flux and a better contrast ratio than shown in Figs. 8 to 14.

*4.1 Earth-like Visible/Near-infrared Spectra (0.4μm – 4μm)*

Fig. 8 shows spectra from 0.4 to 2μm of Earth-like planets for both a clear-sky and Earth-analogue cloud cover for the grid stars (F0V-K7V). The high resolution spectra have been smoothed to a resolving power of 800 using a triangular smoothing kernel. Figs. 8 and 9 show that clouds increase the reflectivity of an Earth-like planet in the VIS to NIR substantially and therefore overall increase the equivalent width of all observable feature, even though they block access to some of the lower atmosphere.

Fig. 9 shows individual features for the strongest atmospheric features from 0.4 to 4μm for Earth-like planets orbiting the grid stars: $O_3$ at 0.6 μm (the Chappuis band), $O_2$ and 0.76 μm, $H_2O$ at 0.95 μm, and $CH_4$ at 1.7μm. The left panel of each row shows the relative flux as planet-to-star contrast ratio, the middle and right panel show the specific, top-of-atmosphere flux for a clear and 60% cloud cover, respectively. From the planet-to-star contrast ratios in Figs. 9, 11 and 13 the photometric precision required to detect these features for Earth-like planets can be calculated. Note that any shallow spectral features like the visible $O_3$ feature would require a very high SNR to be detected.

The 0.6 μm shallow $O_3$ spectral feature depth increases with $T_{eff}$ of the star host since $O_3$ concentration increases with UV levels but is difficult to distinguish from Rayleigh scattering. The relative depth of the $O_2$ feature at 0.76 μm is constant but the flux decreases for cool grid stars due to the decrease in absolute stellar flux received and reflected by the planet at short wavelengths. The depth of the $H_2O$ absorption feature at 0.9 μm (shown) 0.8, 1.1 and 1.4 μm increase for planets orbiting cool grid stars due to their increased $H_2O$ abundance. The depth of the $CH_4$ absorption feature at 1.7μm increases with decreasing stellar $T_{eff}$ due to the increase of $CH_4$ abundance.

From 2 to 4 μm there are $CH_4$ features at 2.3μm and 3.3 μm, a $CO_2$ feature at 2.7μm, and $H_2O$ absorption at 2.7μm and 3.7μm. However, due to the low emergent flux in this region, these features are not shown individually.

*4.2 Earth-like Infrared Spectra, IR (4μm – 20μm)*

Fig. 10 shows spectra from 4 to 20μm of Earth-like planets for both a clear sky and Earth-analogue cloud cover for the grid stars (F0V-K7V). The high resolution spectra have been smoothed to a resolving power of 250 using a triangular smoothing kernel. Clouds decrease the overall emitted flux of an Earth-like planet in the IR.

Fig. 11 shows individual features for the strongest atmospheric features from 4 to 20μm for Earth-like planets orbiting the grid stars: $O_3$ at 9.6μm, $CO_2$ at 15 μm, $H_2O$ at 6.3μm and $CH_4$ at 7.7μm for a cloud free and Earth-analogue cloud coverage model. The left panel of each row shows the relative flux as planet-to-star contrast ratio, the middle and right panel show the specific, top-of-atmosphere flux for a clear and 60% cloud coverage case, respectively.

In the clear sky model, the depth of the $O_3$ feature at 9.6μm decreases for planet models orbiting hot grid stars, despite increasing $O_3$ abundance, due to lower contrast between the continuum and absorption layer temperature. For Earth-analogue cloud cover, however, $O_3$ is seen in emission for $T_{eff} \geq$ 6500K due to the lower continuum temperature.

Due to the hot stratosphere for all grid stars with $T_{eff} > 6000K$, the $CO_2$ absorption feature at 15 μm has a prominent central emission peak. Clouds reduce the continuum level and the depth of the observable $CO_2$ feature.

The $CH_4$ feature at 7.7μm is prominent in the planetary spectra around cool grid stars due to high $CH_4$ abundance in low UV environments. The $CH_4$ feature is also partially obscured by the wings of the $H_2O$ feature at 5-8μm. The depth of the $H_2O$ features at 5-8 and 18+ μm do not change significantly even though $H_2O$ abundance increases for cool grid stars. Clouds reduce the continuum level and the depth of the observable $H_2O$ features.

Fig. 13 shows planet-to-star contrast ratio for Earth-analog cloud cover of an Earth-like planet from which the photometric precision required can be calculated. The planet-to-star contrast ratio is between $10^{-8}$ to $10^{-11}$ in the VIS/NIR and between about $10^{-6}$ and $10^{-10}$ in the IR for the grid stars. For the whole wavelength range, the contrast ratio improves for cool grid stars.

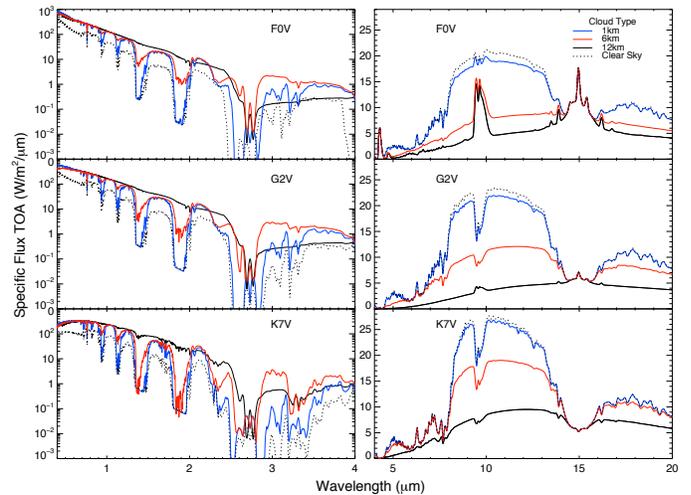

Figure 12: Spectra of Earth-like planets for 100% cloud coverage at 3 cloud heights (1km, 6km and 12km, blue, red and black line, respectively) as well as clear sky spectrum (dashed line) from 0.4 to 20 μm, orbiting a $T_{eff}$ = 7000K (top) $T_{eff}$ = 5750K (middle), and $T_{eff}$ = 4250K (bottom) grid star for comparison.

*4.3 The effect of clouds on an Earth-like planet spectra from 0.4 to 20μm*

Fig. 12 shows Earth-like planet spectra for 100% cloud cover at 1km, 6km and 12km from 0.4 to 20μm for three sample grid stars with $T_{eff}$ = 7000K (top), 5750K (middle), and 4250K (bottom). The clear sky spectrum is shown as dashed line for comparison. Clouds increase the reflectivity of an Earth-like planet in the VIS to NIR substantially and therefore overall increase the equivalent width of all observable features, even though they block access to some of the lower atmosphere. Clouds decrease the overall emitted flux of an Earth-like planet in the IR slightly because they radiate at lower temperatures and therefore overall decrease the equivalent width of all observable absorption features, even though they can increase the relative depth of a spectral feature due to lowering the continuum temperature of the planet.

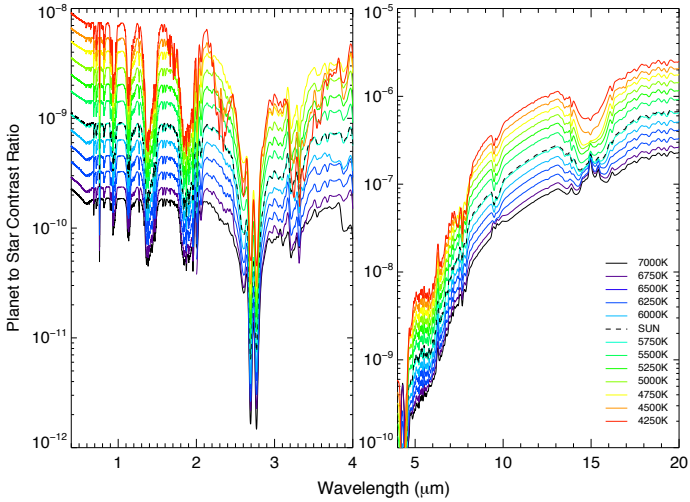

Figure 13: Contrast Ratio of Earth-like planets for Earth-analogue cloud coverage.

Fig. 14 shows the individual chemical absorption features as discussed in section 4.1 and 4.2 on a relative scale for $H_2O$, $CO_2$, $O_2$, $O_3$, $CH_4$, $N_2O$ and $CH_3Cl$ from 0.4μm to 20μm to complement the spectra shown in Figs. 5-9, that focus on the remote detectability of individual features for future space missions.

## 5. DISCUSSION

When choosing IUE stars to for our stellar spectral grid, we avoided stars of unusual variability, but did not exclude stars that had representative variability of its stellar class. Several of our representative K stars are variables of the BY Draconis type which is a common variable in this stellar type. We preferentially choose stars with near solar metallicity when possible; however, the IUE database does not provide candidate stars at each temperature of solar metalliticy. Several stars have lower than solar metallicity. We compared a subsolar stellar metallicity with a solar metallicity spectra model and found that the difference does not impact our results.

*Observability of Biosignatures:* Detecting the combination of $O_2$ or $O_3$ and $CH_4$ for emergent spectra and secondary eclipse measurements requires observations in the IR or in the VIS/NIR up to 3 μm to include the 2.4 μm $CH_4$ feature in that spectral range. The strength of the absorption features depend on the stellar effective temperature of the host star and vary significantly between stellar types. In the IR, $CH_4$ at 7.7 μm is more detectable at low resolution for cool grid stars than hot grid stars. The 9.6 μm $O_3$ feature is deepest for mid to cool stars and becomes less detectable for hotter stars. However around our hottest grid stars, the 9.6 μm $O_3$ feature becomes an apparent emission feature for cloudy atmospheres. The narrow $O_2$ feature in the VIS at 0.72 μm is of comparable strength for all grid stars. $H_2O$ has strong features for all grid stars over the whole wavelength range.

$N_2O$ and $CH_3Cl$ have features from the NIR to IR (see Fig. 14) but in modern Earth concentrations do not have a strong enough feature to be detected with low resolution. For the clear sky models, the vegetation red edge is detectable due to the order of magnitude increased reflectance from 0.7 μm to 0.75 μm for all grid stars. Clouds obscure that feature (see Fig. 8).

For detecting an oxidizing gas in combination with a reducing gas in Earth-like planet atmosphere models, the coolest grid stars in our sample are the best targets. In this paper we have not modeled planets orbiting stars cooler than ~4000K to provide a consistent set of planetary models. As discussed in Segura et al. 2005, cool host stars with low UV flux, provide an environment that leads to run-away $CH_4$ accumulation in the atmosphere and therefore the model for Earth-like planets around M-dwarfs often use abiotic $CH_4$ levels, not consistent with Earth-analogue models used in this study. We will explore this effect in a future work.

No noise has been added to these model spectra to provide input models for a wide variety of instrument simulators for both secondary eclipse and direct detection simulations. Different instrument simulators for JWST (see e.g. Deming *et al.*, 2009, Kaltenegger & Traub 2009) explore the capability of JWST's MIRI and NIRspec Instrument to characterize extrasolar planets down to Earth-like planets, with interesting results for planets around close-by as well as luminous host stars. Several new results are forthcoming by several groups that will provide realistic instrument parameters that can be used to determine detectability of these absorption features. Furture ground and space based telescopes are being designed to characterize exoplanets down to Earth-like planets and will provide interesting opportunities to observe atmospheric features, especially for Super-Earths, with radii up to 2 time Earth's radius and therefore 4 times the flux and planet-to-star contrast ratio levels quoted for Earth-size planets shown in Figs. 8-14.

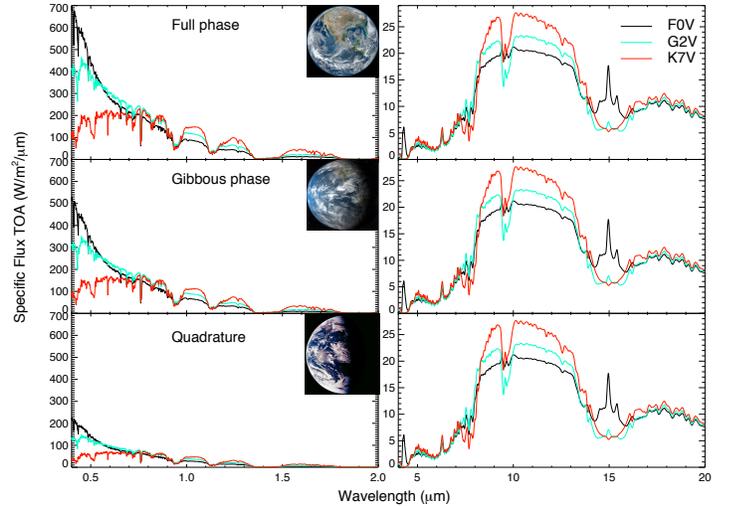

Figure 15: Absolute specific flux values for 60% cloud coverage Earth-like planets around three different grid stars with $T_{eff}$ = 7000K, 5750K, and 4250K in the visible and IR for 3 phases: full phase, gibbous phase, and quadrature with corresponding phase angles of 0°, 45°, and 90°, respectively.

In addition to the size of the planet, future observations will occur at different positions throughout the planet's orbit. The maximum observable planetary flux in the visible scales with the illuminated fraction of the planet, that is "visible" to the observer. In the IR the maximum flux remains constant throughout the planet's orbit, assuming a similar temperature on the day and night side. In Fig. 15 we show the absolute specific flux levels at full phase, gibbous phase, and quadrature (phase angles of 0°, 45°, and 90°, respectively) for 60% cloud coverage Earth-like planets orbiting three grid stars with $T_{eff}$ = 7000K, 5750K and 4250K to show the effect of orbital position (see also Robinson *et al.*, 2011). We scaled our full-phase simulations to other phases using a Lambert phase function. For quadrature, representing an average viewing geometry, the contrast ratios presented in Fig. 13 will be a factor of ~2 lower in the visible. Assuming the planet has efficient heat transport from the day to night side, the specific flux levels and contrast ratios in the IR will be unchanged.

## 6. CONCLUSIONS

We calculated the spectra for terrestrial atmosphere models receiving the same incoming flux as Earth when orbiting a grid of host stars with $T_{eff}$ = 4250K to $T_{eff}$ = 7000K in 250K increments, comprehensively covering the full FGK stellar range. We discuss the spectral features for clear and cloudy atmosphere models and compare the effect of the stars SED and UV flux on both the atmsopheric composition as well as the detectable atmospheric features in section 3 and 4.

Increasing UV environments (generally coupled with increasing stellar $T_{eff}$ for main sequence stars) result in: increasing concentration of $O_3$ from photolysis, increasing stratospheric $H_2O$ from $O_3$ shielding, increasing OH based on increased $O_3$ and $H_2O$ concentrations, and decreasing $CH_4$, $CH_3Cl$, and $N_2O$ from photolysis and reactions with OH. Increasing stellar temperatures and corresponding decreasing planetary surface temperatures result in: decreasing tropospheric $H_2O$ due to decreased temperatures, decreasing stratospheric $H_2O$ from transport, and decreasing reaction rates of OH with $CH_4$, $N_2O$ and $CH_3Cl$. The overall effect as the stellar effective temperature of the main sequence grid stars increases, is an increase in $O_3$ and OH concentration, a decrease in tropospheric $H_2O$ (but an increase stratospheric $H_2O$), and a decrease in stratospheric $CH_4$, $N_2O$, $CH_3Cl$.

In the infrared, the temperature contrast between the surface and the continuum layer is strongly impacts the depth of spectral features. While $O_3$ increases for hotter main sequence stars the strength of the 9.6μm band decreases due to the decrease temperature difference between the continuum and the emitting layer. For hot stars, with $T_{eff} \geq 6750$K the $O_3$ feature appears as emission due to the contrast to the continuum.

Our results provides a grid of atmospheric compositions as well as model spectra from the VIS to the IR for JWST and other future direct detection mission design concepts. The model spectra in this paper are available at www.cfa.harvard.edu/~srugheimer/FGKspectra/.

**Acknowledgements:** L.K. acknowledge support from DFG funding ENP Ka 3142/1-1 and NAI. This research has made use of the NASA/IPAC/NExScI Star and Exoplanet Database, which is operated by the Jet Propulsion Laboratory, California Institute of Technology, under contract with the National Aeronautics and Space Administration.

Some of the data presented in this paper were obtained from the Multimission Archive at the Space Telescope Science Institute (MAST). STScI is operated by the Association of Universities for Research in Astronomy, Inc., under NASA contract NAS5-26555. Support for MAST for non-HST data is provided by the NASA Office of Space Science via grant NAG5-7584 and by other grants and contracts.

**Author Disclosure Statement:** No competing financial interests exist.

Figure 14: Relative absorption of individual chemical species $H_2O$, $CO_2$, $O_2$, $O_3$, $CH_4$, $N_2O$ and $CH_3Cl$ for three sample grid stars with $T_{eff}$ = 7000K, 5750K, and 4250K.

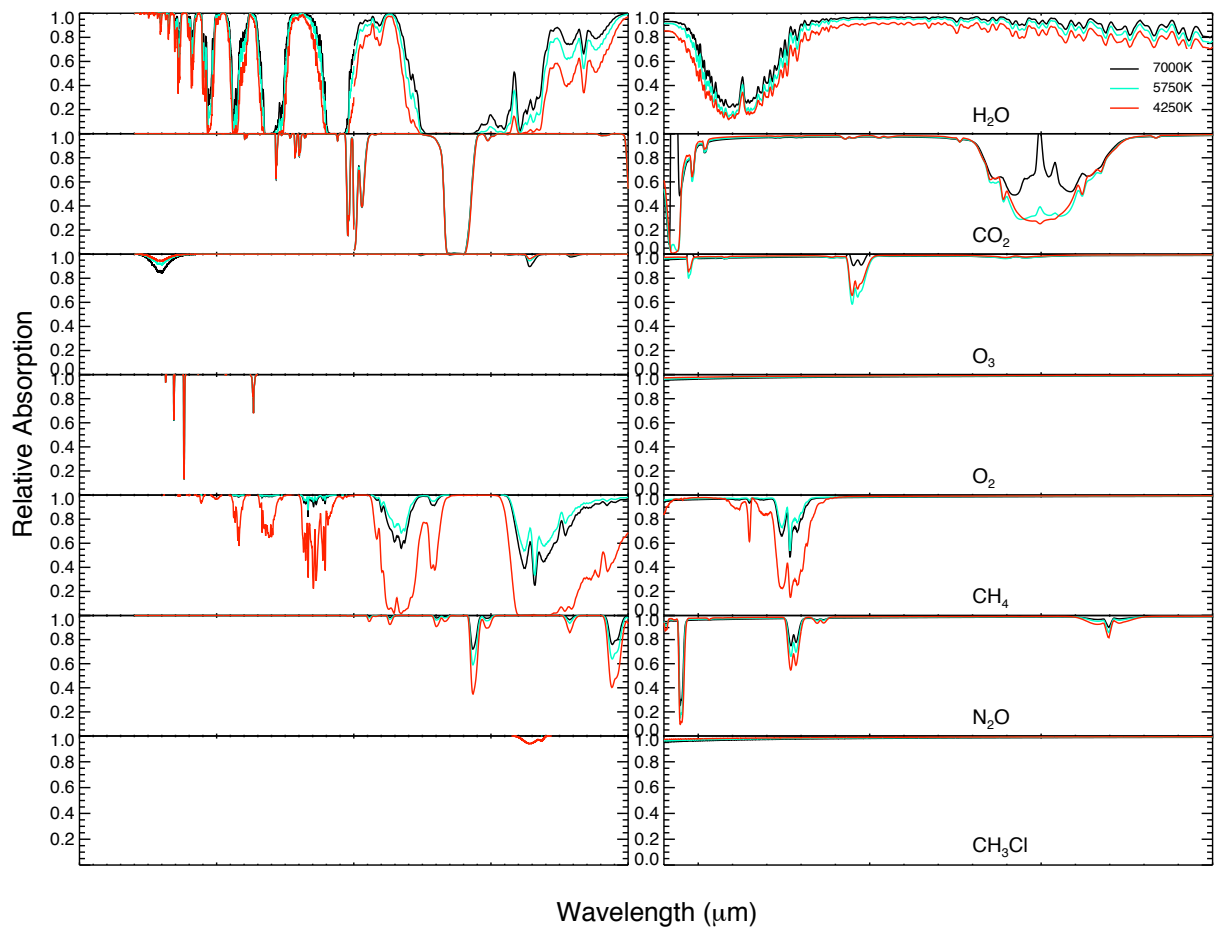